# *Spontaneous emission of an atom in a uniform and nonuniform spaces*

V.S.Zuev

The P.N.Lebedev Physical Institute of RAS
Email: vizuev@sci.lebedev.ru

On purpose to establish the differences in probabilities of spontaneous radiative emission of an atom in a free space and in a nonuniform space with a nanospheroid the field strength $\vec{E}$ and the density of states $\rho_E$ of the mode $\vec{n}_{10}$ of a free space and of the $TM_0$ nanocylinder mode are defined. A nanocylinder is an approximate model of a nanospheroid. The probability of a spontaneous transition is proportional to $\vec{E}^2 \rho_E$ in the both cases. It is found that the emission probability to the nanospheroid mode is hundreds times and more higher than the spontaneous emission probability in the free space.

# *Спонтанное излучение атома в однородном и неоднородном пространствах*

В.С.Зуев

Физический ин-т им. П.Н.Лебедева РАН
Email: vizuev@sci.lebedev.ru

С целью выяснения соотношения между вероятностями излучательного спонтанного перехода в свободном пространстве и в неоднородном пространстве с наносфероидом рассчитаны поле $\vec{E}$ и плотность состояний $\rho_E$ в моде $\vec{n}_{10}$ свободного пространства и в моде $TM_0$ наноцилиндра. Наноцилиндр приближенно моделирует наносфероид. Вероятность спонтанного перехода в каждом из 2-х случаев пропорциональна $\vec{E}^2 \rho_E$. Излучение в моду наносфероида на многие порядки превышает вероятность спонтанного излучения в свободном пространстве.

## *Спонтанное излучение атома в однородном и неоднородном пространствах*

В.С.Зуев

Физический ин-т им. П.Н.Лебедева РАН
Email: vizuev@sci.lebedev.ru

Данный текст написан в сентябре 2006 г., но по случайной причине остался опубликованным лишь в виде препринта ФИАН № 26, 2006г. Направляя текст в печать, исправляем допущенную оплошность.

### *Атом, излучающий в поверхностный плазмон наносфероида*

Поле наносфероида рассмотрим приближенно, так, как это сделано в работе /1/. Наносфероид будет заменен наноцилиндром. Среди мод цилиндра при малом диаметре актуальной является мода $\vec{n}_{0\gamma} \cdot e^{ihz}$, существующая на цилиндре любого диаметра, как бы мал этот диаметр не был. Эта мода имеет нулевую критическую частоту. Эту моду будем называть поверхностной $TM_0$ волной ($TM_0$ плазмоном). Магнитное поле $TM_0$ волны имеет только одну поперечную $\theta$ - компоненту. Функция $\vec{n}_{0\gamma}$ имеет вид /2/:

$$\vec{n}_{0\lambda} = \frac{ih}{\sqrt{k^2}}\frac{d}{dr}Z_0(\lambda r)\cdot\vec{i}_r + \frac{\lambda^2}{\sqrt{k^2}}Z_0(\lambda r)\cdot\vec{i}_z. \tag{1}$$

Формула (1) описывает поле как внутри цилиндра, среда $i = 1$, так и вне его, среда $i = 2$, но в разных средах содержит разные цилиндрические функции и две различные пары значений $k^2$ и $\gamma$, а именно $k_1^2$, $\gamma_1^2 = k_1^2 - h^2$ и $k_2^2$, $\gamma_2^2 = k_2^2 - h^2$. $TM_0$ волна – неоднородная волна, $k_1^2 < 0$, $0 < k_2^2 <|k_1^2|$, $h^2 > k_1^2, k_2^2$, $\gamma_1^2, \gamma_2^2 < 0$. Внутри цилиндра вместо $Z_0(\gamma r)$ следует выбрать $J_0(\gamma_1 r)$, вне цилиндра - $H_0^{(1)}(\gamma_2 r)$. Эти функции имеют мнимый аргумент, что позднее учтем и перейдем к модифицированным бесселевым функциям.

Зададим электрическую напряженность $\vec{E}$, а по уравнению $\nabla\times\vec{E} - i\frac{\mu\omega}{c}\vec{H} = 0$ определим соответствующее $\vec{H}$. Используем соотношение $\frac{dZ_0(\rho)}{d\rho} = -Z_1(\rho)$.

$$\vec{E} = E\vec{n}_{0\lambda}e^{ihz} = E\left[-\frac{ih\lambda}{\sqrt{k^2}}Z_1(\lambda r)\cdot\vec{i}_r + \frac{\lambda^2}{\sqrt{k^2}}Z_0(\lambda r)\cdot\vec{i}_z\right]e^{ihz}, \tag{2}$$

$$\vec{H} = -i\frac{c}{\omega\mu}\nabla\times\vec{E}. \tag{3}$$

В цилиндрических координатах $r, \theta, z$ имеем

$$\nabla\times\vec{F} = \left(\frac{1}{r}\frac{\partial F_z}{\partial\theta} - \frac{\partial F_\theta}{\partial z}\right)\cdot\vec{i}_r + \left(\frac{\partial F_r}{\partial z} - \frac{\partial F_z}{\partial r}\right)\cdot\vec{i}_\theta + \left[\frac{1}{r}\frac{\partial}{\partial r}(rF_\theta) - \frac{1}{r}\frac{\partial F_r}{\partial\theta}\right]\cdot\vec{i}_z. \tag{4}$$

Для поля вида (2) с отличными от нуля компонентами $E_r$ и $E_z$, независящими от $\theta$, и $E_\theta = 0$ получим

$$\nabla \times \vec{E} = \left(\frac{\partial E_r}{\partial z} - \frac{\partial E_z}{\partial r}\right) \cdot \vec{i}_\theta, \ (\nabla \times \vec{E})_\theta = -E\left(\frac{h^2 \lambda}{\sqrt{k^2}} + \frac{\lambda^3}{\sqrt{k^2}}\right) Z_1(\lambda r) e^{ihz} \tag{5}$$

Итак, поля имеют вид

$$E_r = \begin{cases} E_1 \frac{h\gamma_1}{\kappa_1} I_1(\gamma_1 r) e^{ihz} \\ -E_2 \frac{h\gamma_2}{k_2} K_1(\gamma_2 r) e^{ihz} \end{cases}, \ E_z = \begin{cases} E_1 \frac{i\gamma_1^2}{\kappa_1} I_0(\gamma_1 r) e^{ihz} \\ E_2 \frac{i\gamma_2^2}{k_2} K_0(\gamma_2 r) e^{ihz} \end{cases}, \ H_\theta = \begin{cases} \gamma_1 \sqrt{\frac{|\varepsilon_1|}{\mu_1}} E_1 I_1(\gamma_1 r) e^{ihz} \\ \gamma_2 \sqrt{\frac{\varepsilon_2}{\mu_2}} E_2 K_1(\gamma_2 r) e^{ihz} \end{cases}. \tag{6}$$

Внутри цилиндра - среда $i=1$, вне - среда $i=2$. $\varepsilon_1 < 0$, $h^2 + \lambda_i^2 = k_i^2$. Обе $\lambda_i^2 < 0$, поэтому введены $\gamma_i^2 = -\lambda_i^2$, $\kappa_1^2 = -k_1^2$. В (6) произведена замена цилиндрических функций на модифицированные функции по формулам (7). Множитель $2/\pi$ включен в $E_2$.

$$I_n(|\gamma|r) = \exp(-in\pi/2)J_n(i|\gamma|r), \ J_0(i|\gamma|r) = I_0(|\gamma|r), \ J_1(i|\gamma|r) = iI_1(|\gamma|r).$$
$$K_n(|\gamma|r) = (i\pi/2)\exp(in/\pi)H_n^{(1)}(i|\gamma|r), \ H_0^{(1)}(i|\gamma|r) = -i(2/\pi)K_0(|\gamma|r), \tag{7}$$
$$H_1^{(1)}(i|\gamma|r) = -(2/\pi)K_1(|\gamma|r).$$

Теперь составляем уравнения граничных условий равенства тангенциальных компонент полей, составляем детерминант полученной системы и в итоге получаем характеристичекое уравнение:

$$\frac{\gamma_1}{\gamma_2} \frac{\varepsilon_2}{|\varepsilon_1|} \frac{K_1(\gamma_2 a)}{K_0(\gamma_2 a)} = \frac{I_1(\gamma_1 a)}{I_0(\gamma_1 a)}. \tag{8}$$

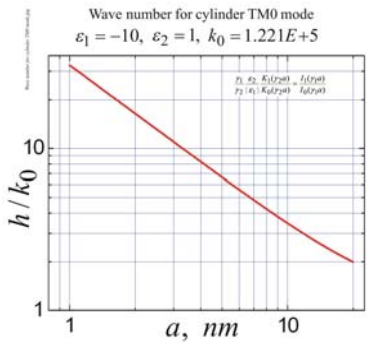

Рис.1.

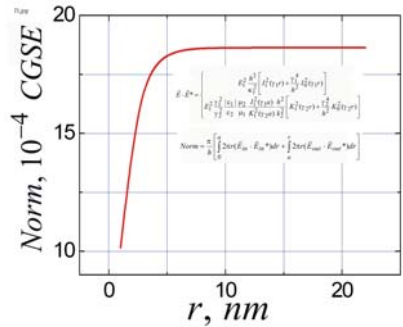

Рис.2.

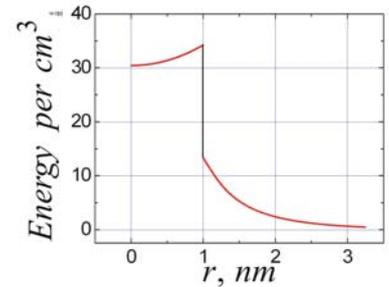

Рис.3. Плотность энергии в зависимости от радиуса для наноцилиндра радиуса 1 нм.

Определяем связь между амплитудами:

$$E_2 = E_1 \frac{\gamma_1^2}{\gamma_2^2} \sqrt{\frac{\varepsilon_2 \mu_2}{|\varepsilon_1|\mu_1}} \frac{I_0(\gamma_1 a)}{K_0(\gamma_2 a)}. \tag{9}$$

Отношение амплитуд:

| 1 | 2 | 4 | 6 | 8 | 10 | 20 |
|---|---|---|---|---|----|----|
| 0.298 | 0.308 | 0.346 | 0.411 | 0.502 | 0.621 | 1.691 |

Теперь пронормируем поле, приравняв энергию поля в моде одному фотону. Мгновенное значение плотности энергии э.-м. поля в плазме выражается формулой

$$w = \frac{1}{8\pi}(\vec{E}\cdot\vec{E}) + \frac{1}{8\pi}(\vec{H}\cdot\vec{H}) + \frac{m_e v_e^2}{2} n_e \qquad (10)$$

В формулу для энергии (10) поля должны быть взяты в вещественной форме.

$$E_{1r} = E_1 \frac{h\gamma_1}{\kappa_1} I_1(\gamma_1 r)\cos(hz-\omega t), \quad E_{1z} = -E_1 \frac{\gamma_1^2}{\kappa_1} I_0(\gamma_1 r)\sin(hz-\omega t), \quad H_{1\theta} = \gamma_1 \sqrt{\frac{|\varepsilon_1|}{\mu_1}} E_1 I_1(\gamma_1 r)\cos(hz-\omega t).$$
$$E_{2r} = -E_2 \frac{h\gamma_2}{k_2} K_1(\gamma_2 r)\cos(hz-\omega t) \quad E_{2z} = -E_2 \frac{\gamma_2^2}{k_2} K_0(\gamma_2 r)\sin(hz-\omega t) \quad H_{2\theta} = \gamma_2 \sqrt{\frac{\varepsilon_2}{\mu_2}} E_2 K_1(\gamma_2 r)\cos(hz-\omega t)$$
$$(11)$$

Энергию единицы объема плазмы рекомендуют брать в виде $(E^2+H^2)/8\pi + n_e m_e v_e^2/2$. Плотность электронов $n_e$ определим из $\varepsilon_{pl} = 1 - \omega_{pl}^2/\omega^2$ для плазмы, где $\omega_{pl}^2 = 4\pi e^2 n_e / m_e$. Энергия электрона равна $m_e v_e^2/2$, $m_e dv_e/dt = eE_0 e^{-i\omega t}$, $v_e = -(eE_0/\omega m_e)\sin\omega t$, $\frac{m_e v_e^2}{2} = \frac{m_e}{2}\frac{e^2 E_0^2}{\omega^2 m_e^2}\sin^2\omega t$.

Энергия электронов в единице объема равна $w_e = \frac{1}{8\pi}\frac{\omega_{pl}^2}{\omega^2} E_0^2 \sin^2\omega t = \frac{1}{8\pi}(1-\varepsilon_{pl})E_0^2\sin^2\omega t$. Таким образом энергия тела с отрицательным $\varepsilon$ (плазма) равна $w = \frac{1}{8\pi}[(2-\varepsilon)E^2 + \mu H^2]$, хотя, казалось бы, она должна быть равна $w = \frac{1}{8\pi}(-\varepsilon E^2 + \mu H^2)$. При большом $|\varepsilon|$ разница в формулах не велика.

Теперь вычисляем квадраты полей.

$$(\vec{E}_1 \cdot \vec{E}_1) = E_1^2 \frac{\gamma_1^2}{\kappa_1^2}\left[h^2 I_1^2(\gamma_1 r)\cos^2(hz-\omega t) + \gamma_1^2 I_0^2(\gamma_1 r)\sin^2(hz-\omega t)\right],$$
$$(\vec{E}_2 \cdot \vec{E}_2) = E_2^2 \frac{\gamma_2^2}{k_2^2}\left[h^2 K_1^2(\gamma_2 r)\cos^2(hz-\omega t) + \gamma_2^2 K_0^2(\gamma_2 r)\sin^2(hz-\omega t)\right], \qquad (12)$$
$$(\vec{H}_1 \cdot \vec{H}_1) = \gamma_1^2 \frac{|\varepsilon_1|}{\mu_1} E_1^2 I_1^2(\gamma_1 r)\cos^2(hz-\omega t),$$
$$(\vec{H}_2 \cdot \vec{H}_2) = \gamma_2^2 \frac{\varepsilon_2}{\mu_2} E_2^2 K_1^2(\gamma_2 r)\cos^2(hz-\omega t),$$

$$w_1 = [(2-\varepsilon_1)E_1^2 + \mu_1 H_1^2]/8\pi, \quad w_2 = (\varepsilon_2 E_2^2 + \mu_2 H_2^2)/8\pi. \qquad (13)$$

$$w_1 = \frac{E_1^2}{8\pi}(2+|\varepsilon_1|)\frac{\gamma_1^2}{\kappa_1^2}\left[\left(h^2 + \frac{\kappa_1^2 |\varepsilon_1|}{(2+|\varepsilon_1|)}\right) I_1^2(\gamma_1 r)\cos^2(hz-\omega t) + \gamma_1^2 I_0^2(\gamma_1 r)\sin^2(hz-\omega t)\right],$$
$$w_2 = \frac{E_1^2}{8\pi}\varepsilon_2 \frac{\gamma_1^2}{\kappa_1^2}\left[\frac{\gamma_1^2}{\gamma_2^2}\frac{I_0^2(\gamma_1 a)}{K_0^2(\gamma_2 a)}\right][(h^2+k_2^2)K_1^2(\gamma_2 r)\cos^2(hz-\omega t) + \gamma_2^2 K_0^2(\gamma_2 r)\sin^2(hz-\omega t)]. \qquad (14)$$

Формулы для расчета, $A^2 = \frac{E_1^2}{8\pi}\frac{\gamma_1^2}{\kappa_1^2}\frac{\pi}{2h}$, $\int_0^\pi \cos^2 x\, dx = \int_0^\pi \sin^2 x\, dx = \pi/2$:

$$\overline{w}_1 = (2+|\varepsilon_1|)A^2\left[\left(h^2 + \frac{\kappa_1^2|\varepsilon_1|}{(2+|\varepsilon_1|)}\right)I_1^2(\gamma_1 r) + \gamma_1^2 I_0^2(\gamma_1 r)\right],$$
$$(15)$$

$$\overline{w}_2 = \varepsilon_2 A^2 \left[ \frac{\gamma_1^2}{\gamma_2^2} \frac{I_0^2(\gamma_1 a)}{K_0^2(\gamma_2 a)} \right] [(h^2 + k_2^2) K_1^2(\gamma_2 r) + \gamma_2^2 K_0^2(\gamma_2 r)].$$

Отношение энергии в наноцилиндре к энергии вне наноцилиндра равно приблизительно 1.2 для $a = 1 \div 5 \ nm$ и для $\varepsilon_2 = 1 \div 9$.

$$W = 2\pi A^2 \left( \int_0^a r\overline{w}_1 dr + \int_a^\infty r\overline{w}_2 dr \right)_{a=1\,nm} = 11.864 A^2 = 11.864 \frac{E_1^2}{8\pi} \frac{\gamma_1^2}{\kappa_1^2} \frac{\pi}{2h}. \quad (16)$$

В итоге получаем:

$$E_1^2 = \frac{16h}{11.864} \frac{\kappa_1^2}{\gamma_1^2} \hbar\omega. \quad (16')$$

Плотность состояний для данной моды /1/:

$$\rho_E = \frac{Q}{\hbar\omega}. \quad (16'')$$

Теперь вычислим поле в зазоре $\vec{E}_{gap}$. Поскольку

$$E_{1z} = -E_1 \frac{\gamma_1^2}{\kappa_1} I_0(\gamma_1 r) \sin(hz - \omega t),$$

то

$$\vec{E}_{gap} = -\varepsilon_1 E_1 \frac{\gamma_1^2}{\kappa_1} I_0(\gamma_1 r = 0) \sin[(hz = h\lambda_{pl}/2) - \omega t] = -\varepsilon_1 E_1 \frac{\gamma_1^2}{\kappa_1}.$$

### *Спонтанное излучение атома в свободном пространстве с излучением в сферические моды*

Спонтанное излучение атома в свободном пространстве с излучением в сферические моды было рассмотрено в работе /3/. Отметим, что представление поля в свободном пространстве в виде набора плоских волн или набора сферических волн не меняет конечного результата, а именно, величины вычисляемой вероятности спонтанного перехода.

Если начало системы $r, \theta, \varphi$ находится в точке расположения атома и полярная ось совпадает с $\vec{p}$, импульсом оптического электрона в атоме, то среди всех мод актуальной является единственная мода $\vec{n}_{10}$. Эта мода выглядит так /2/:

$$\vec{n}_{10} = 2P_1(\cos\theta) \frac{j_1(\rho)}{\rho} \cdot \vec{i}_r + \frac{d}{d\theta} P_1(\cos\theta) \frac{1}{\rho} \frac{d}{d\rho}[\rho j_1(\rho)] \cdot \vec{i}_\theta. \quad (17)$$

С учетом формул

$$P_1(\cos\theta) = \cos\theta, \ \frac{1}{\rho} \frac{d}{d\rho}[\rho j_1(\rho)] = \frac{1}{3}[2j_0(\rho) - j_2(\rho)], \quad (18)$$

$$\frac{1}{\rho} j_1(\rho) = \frac{1}{3}[j_0(\rho) + j_2(\rho)], \quad j_2 = \frac{3}{\rho} j_1 - j_0.$$

получаем

$$\vec{n}_{10} = \frac{2}{3}(j_0 + j_2)\cos\theta \cdot \vec{i}_r - \frac{1}{3}(2j_0 - j_2)\sin\theta \cdot \vec{i}_\theta,$$
$$\vec{n}_{10} = \frac{2}{\rho} j_1 \cos\theta \cdot \vec{i}_r - (j_0 - \frac{1}{\rho} j_1)\sin\theta \cdot \vec{i}_\theta. \qquad (19)$$

Зададим электрическую напряженность $\vec{E}$

$$\vec{E} = E\vec{n}_{10} = E\left[\frac{2}{\rho} j_1 \cos\theta \cdot \vec{i}_r - (j_0 - \frac{1}{\rho} j_1)\sin\theta \cdot \vec{i}_\theta\right], \qquad (20)$$

а по уравнению $\nabla \times \vec{E} - i\frac{\mu\omega}{c}\vec{H} = 0$ определим соответствующее $\vec{H}$. Вместо прямого вычисления $\nabla \times \vec{E}$ воспользуемся готовым результатом, а именно связью между функциями $\vec{n}_{10}$ и $\vec{m}_{10}$ /2/:

$$\vec{m}_{10} = \frac{1}{k}\nabla \times \vec{n}_{10}, \quad \vec{H} = -i\frac{c}{\mu\omega}\nabla \times \vec{E} = -iE\sqrt{\frac{\varepsilon}{\mu}}\frac{1}{k}\nabla \times \vec{n}_{10} = -iE\sqrt{\frac{\varepsilon}{\mu}}\vec{m}_{10}. \qquad (21)$$

$$\vec{m}_{10} = -j_1(\rho)\frac{d}{d\theta}P_1(\cos\theta) \cdot \vec{i}_\varphi = j_1(\rho)\sin\theta \cdot \vec{i}_\varphi, \quad \vec{H} = -iE\sqrt{\frac{\varepsilon}{\mu}} j_1(\rho)\sin\theta \cdot \vec{i}_\varphi. \qquad (22)$$

С учетом временного множителя поля имеют вид:

$$\vec{E}(t) = E\left\{\frac{2}{\rho} j_1(\rho)\cos\theta \cdot \vec{i}_r - \left[j_0(\rho) - \frac{1}{\rho} j_1(\rho)\right]\sin\theta \cdot \vec{i}_\theta\right\}e^{-i\omega t},$$
$$\vec{H}(t) = -iE\sqrt{\frac{\varepsilon}{\mu}} j_1(\rho)\sin\theta e^{i\omega t} \cdot \vec{i}_\varphi. \qquad (23)$$

В действительной форме поля имеют вид:

$$\vec{E}(t) = E\left\{\frac{2}{\rho} j_1(\rho)\cos\theta \cdot \vec{i}_r - \left[j_0(\rho) - \frac{1}{\rho} j_1(\rho)\right]\sin\theta \cdot \vec{i}_\theta\right\}\cos\omega t,$$
$$\vec{H}(t) = -E\sqrt{\frac{\varepsilon}{\mu}} j_1(\rho)\sin\theta \sin\omega t \cdot \vec{i}_\varphi. \qquad (24)$$

Выбираем такой момент времени, когда $\cos\omega t = 0$, $\sin\omega t = 1$. Мгновенная плотность энергии поля равна

$$w = \varepsilon\frac{(\vec{E})^2}{8\pi} + \mu\frac{(\vec{H})^2}{8\pi} = \frac{\varepsilon E^2 j_1^2(\rho)\sin^2\theta}{8\pi}. \qquad (25)$$

Полная энергия в моде

$$W = 2\pi \int_0^\pi \sin\theta d\theta \int_0^\infty r^2 w(r,\theta) dr = \frac{\varepsilon}{4} E^2 \int_0^\pi (1-\cos^2\theta)\sin\theta d\theta \int_0^\infty r^2 j_1^2(\rho) dr =$$

$$= \frac{\varepsilon}{4} E^2 \int_0^1 (1-x^2) dx \int_0^\infty r^2 j_1^2(kr) dr \underset{kr \gg 1}{\approx} \frac{\varepsilon}{3} E^2 \int_0^R \frac{\rho^2}{k^3} \frac{\cos^2\rho}{\rho^2} d\rho = \frac{\varepsilon E^2 R}{6k^2}$$

$$W = \frac{\varepsilon E^2}{8\pi} \int_V (\vec{n}_{10})^2 dV = \frac{\varepsilon E^2 R}{6k^2}. \tag{26}$$

$$E^2 = \frac{6k^2}{\varepsilon R} \hbar\omega. \tag{27}$$

Плотность состояний /3/:

$$\rho_E = \frac{R}{2\pi\hbar c} \tag{28}$$

## *Отношение вероятности излучения в плазмон к вероятности излучения в пустое пространство*

Теперь составим отношение вероятности излучения в плазмон к вероятности излучения в пустое пространство. Атом находится в щели. Это отношение равно:

$$F = \frac{\dfrac{16h}{11.864}\dfrac{\kappa_1^2}{\gamma_1^2}\hbar\omega\dfrac{Q}{\hbar\omega}}{\dfrac{6k_0^2}{\varepsilon_2 R}\hbar\omega\dfrac{R}{2\pi\hbar c}}\left(\varepsilon_1\dfrac{\gamma_1^2}{\kappa_1}\right)^2 = \frac{\dfrac{16h}{11.864}\dfrac{\kappa_1^2}{\gamma_1^2}\dfrac{Q}{\hbar\omega}}{\dfrac{6k_0^2}{2\pi\hbar c \varepsilon_2}}\left(\varepsilon_1\dfrac{\gamma_1^2}{\kappa_1}\right)^2 \approx \frac{4}{3}\frac{h\gamma_1^2}{k_0^3}\varepsilon_1^2 Q \tag{29}$$

$$h\gamma_1^2 \approx h^3, \quad Q = 70, \quad \varepsilon_1^2 = 10^2, \quad h^3/k_0^3 = (32.8)^3 = 3\cdot 10^4$$

$$\frac{4}{3}\frac{h\gamma_1^2}{k_0^3}\varepsilon_1^2 Q \approx \frac{4}{3} 3\cdot 10^4 \cdot 70 \cdot 10^2 \approx 3\cdot 10^8. \tag{30}$$

Полученное значение отношения чрезвычайно велико. Эту оценку следует рассматривать как качественную, но правильную по порядку величины. Для точного определения эффекта задачу следует решать, не прибегая к методу теории возмущений.